\documentclass[10pt,journal,cspaper,compsoc]{IEEEtran}

\usepackage[english]{babel}
\usepackage{graphicx}
\usepackage{wrapfig}
\usepackage{url}
\usepackage{picinpar}\usepackage{amsmath}
\usepackage{amssymb} 
\usepackage{setspace}
\usepackage{epstopdf}
\usepackage{fancyhdr}

\makeatletter
\newif\if@restonecol
\makeatother

\usepackage[ruled]{algorithm2e}

\newcounter{definitionsCounter}
\newcommand{\CommunitySuperNode}{LeafSuperNode}
\newcommand{\SuperNode}{SuperNode}

\newcommand{\Hgraph}{ {CP}} 
\newcommand{\CEPS}{{\em{CEPS}}}       
\newcommand{\CEPSfull}{{{center-piece}}}       
\newcommand{\pmetric}{\emph{meeting probability}}
\newcommand{\ssp}{\emph{steady-state probability}}
\newcommand{\extract}{\emph{EXTRACT}}
\newcommand{\lqu}{``}
\newcommand{\rqu}{''}
\newcommand{\qi}{q_i}  
\newcommand{\pd}{pd}  
\newcommand{\QN}{{Q}}       
\newcommand{\QMySet}{{\cal Q}}    
\newcommand{\collision}{r}  
\newcommand{\cij}{ \collision_{i,j}}    
\newcommand{\cijpar}{ \collision{(i,j)}} 
\newcommand{\cqj}{ \collision ( \QMySet, j)} 
\newcommand{\cqjQ}{ \collision ( \QMySet, j, \QN)} 
\newcommand{\mat}[1]{{\bf #1}}   
\newcommand{\matC}{ \mat{R}} 
\newcommand{\matW}{G'} 
\newcommand{\matWnorm}{\bf\tilde{G'}} 
\newcommand{\unitvi}{ \vec{e}_i} 
\newcommand{\bit}{\begin{itemize}}
\newcommand{\eit}{\end{itemize}}
\newcommand{\hide}[1]{}

\ifCLASSOPTIONcompsoc
   \usepackage[nocompress]{cite}
\else
   \usepackage{cite}
\fi

\begin{document}

\title{Large Graph Analysis in the GMine System\\
\ \ \ \ \ \ \ \ \ \ \ \ \ \ \ \ \ \ \ \ \ \ \ \ \ \ \ \ \ \ \ \ \ \ \ \ \ \ \ \ \ \ \ \ \ \ \ \ \ \ \ \  \ \ \ \ \ \ \ \small{http://dx.doi.org/10.1109/TKDE.2011.199} -- http://ieeexplore.ieee.org/xpl/articleDetails.jsp?arnumber=6025354}

\author{Jose F. Rodrigues Jr., $\dagger$ Hanghang Tong, +Jia-Yu Pan, Agma J. M. Traina, Caetano Traina Jr., *Christos Faloutsos
\thanks{Inst. de Ci\^encias Matem\'aticas e de Computa\c{c}\~ao - Universidade de S\~ao Paulo - 13560-970 S\~ao Carlos, SP, Brazil - \{junio, agma, caetano\}@icmc.usp.br}
\thanks{$\dagger$ IBM T.J. Watson Research - 19 Skyline Dr. - Hawthorne NY, 10532 - htong@us.ibm.com}
\thanks{*School of Computer Science - Carnegie Mellon University - 5000 Forbes Ave - 15213-3891, USA - christos@cs.cmu.edu}
\thanks{+Google Inc., Pittsburgh - 4720 Forbes Ave., Lower Level - Pittsburgh, PA 15213 - jiayu.pan@gmail.com}
}

\IEEEcompsoctitleabstractindextext{%
\begin{abstract}
\noindent{Current applications have produced graphs on the order of hundreds of thousands of nodes and millions of edges. To take advantage of such graphs, one must be able to find patterns, outliers and communities. These tasks are better performed in an interactive environment, where human expertise can guide the process. For large graphs, though, there are some challenges: the excessive processing requirements are prohibitive, and drawing hundred-thousand nodes results in cluttered images hard to comprehend. To cope with these problems, we propose an innovative framework suited for any kind of tree-like graph visual design. GMine integrates (a) a representation for graphs organized as hierarchies of partitions - the concepts of SuperGraph and Graph-Tree; and (b) a graph summarization methodology - CEPS. Our graph representation deals with the problem of tracing the connection aspects of a graph hierarchy with sub linear complexity, allowing one to grasp the neighborhood of a single node or of a group of nodes in a single click. As a proof of concept, the visual environment of GMine is instantiated as a system in which large graphs can be investigated globally and locally.} 

\end{abstract}

\begin{keywords}
Graph Analysis System, Graph Representation, Data Structures, Graph Mining, Graph Visualization\\

IEEE Copyright - http://ieeexplore.ieee.org/xpl/articleDetails.jsp?arnumber=6025354
\end{keywords}}

\maketitle

\sloppy
\section{Introduction}
\label{sec:Introduction}

\noindent{Large graphs are common in real-life settings: web graphs, computer communication graphs, recommendation systems, social networks, bipartite graphs of web-logs, to name a few. To find patterns in a large graph, it is desirable to compute, visualize, interact and mine it. However, dealing with graphs on the order of hundreds of thousands of nodes and millions of edges brings some problems: the excessive processing requirements are prohibitive, and drawing hundred-thousand nodes results in cluttered images that are hard to comprehend.

\indent{In former works, the large graph problem has been treated through graph hierarchies, according to which a graph is recursively broken to define a tree of sets of partitions. However, previous efforts on this matter fail on the task of integrating the information from multiple partitions, disregarding mining techniques to fine inspect each subgraph. Conversely, for understanding a graph hierarchy, it is worthwhile to have systems that provide aids for answering the following questions:}

\begin{itemize}
 \item{Hierarchical navigation: {\it What is the relation between arbitrary groups (partitions) of nodes?}}
 \item{Representation and processing: {\it What are the adjacencies of a given graph node considering the entire graph, and not only its particular partition?}}
 \item{Mining: {\it Given a subset of nodes in the graph, what is the induced subgraph that best summarizes the relationships of this subset?}}
\item{Visualization: {\it How do we see through the levels of the graph hierarchy?}} 
 \item{Interaction: {\it How do we perform all these tasks efficiently and intuitively?}}
\end{itemize}

It is our contention that a system that presents the original graph concomitant to its hierarchical version must meet all these requirements. Therefore, we seek for a new representation for graph hierarchies, different from previous works in which the graph hierarchy is ``stagnant'' and cannot answer questions about the relationships between nodes at different groups, and neither between groups  at different partitions of the hierarchy. These are serious limitations because a graph is, essentially, a model for representing relationships.

Another concern is that even at the deepest level of a graph hierarchy -- at the leaves, it is possible to find subgraphs complex enough to surpass the analytical capacity. In this situation, one should be able to summarize the subgraph achieving a small, yet representative, fraction of it; an operation that answers for a deeper level of insight over hierarchical partitionings.

The contribution of this work is the integration of methodologies that address the problems discussed above. We introduce a novel representation for graph hierarchies that extends those of previous works, leading to a model more suitable for presentation and computation. Our methodology also counts on the possibility of graph summarization at the subgraphs (leaves) in a graph hierarchy. The result of our efforts is GMine \cite{RodriguesGMine06}, a system that allows browsing and mining of large graphs in a rich visual environment \cite{RodriguesSuperGraph06}.

The paper is organized as follows: section \ref{sec:RelatedWork} reviews related works for this paper. Section \ref{sec:Terminology} introduces the SuperGraph/Graph-Tree methodology and section \ref{sec:CePs} explains the {\CEPS} graph summarization. Section \ref{sec:Performance} presents experiments on the Graph-Tree performance and section \ref{sec:ceps_accuracy} presents accuracy measures for {\CEPS}. As a proof of concept, section \ref{sec:SuperGraph_Visualization_in_GMine} demonstrates the GMine system. Section \ref{sec:Conclusions} concludes the paper.

\vspace{-0.2mm}
\section{Related Work}
\label{sec:RelatedWork}

\noindent{The interest on large graph analysis has increased in the recent years. This research area includes pattern mining \cite{Melo2010}, influence propagation \cite{Rodriguez2010} and community mining \cite{Fortunato2010}, among others. Such themes can benefit from tools that enable the visual inspection of large graphs.\\

\noindent{\bf Graph Hierarchical Presentation}\\
\noindent{Although many works implicitly define the hierarchical clustering of graphs -- as in the work of Eades and Feng \cite{Eades1997}, most of them do not touch the issue of how such arrangements deal with scalability and processing by means of a well-defined data structure. Batagelj {\it et al.}\cite{Batagelj2010}, for instance, generalizes on the concept of {\it X-graph of Y-graphs} to define a properties-oriented hierarchical clustering of graphs not providing details nor performance evaluation of the implicit data arrangement that supports their processing. Archambault {\it et. al} \cite{Archambault2011} define an ingenious dynamic modification of the graph hierarchy in light of a single node of interest; their system requires the user to reset her/his referential {\it locus} at every new choice of a node with a strictly linear complexity on the basis of seconds delay. Gansner {\it et a.} \cite{Gansner2005} present a fish-eye visualization built over a graph layout with pre-computed coordinates, their structure permits the inspection of the graph at multiple levels of details. Schaffer {\it et al.} \cite{Schaffer1996} describe an earlier fish-eye approach focused on the interactive experience. From the aesthetic perspective, Ham and Wijk \cite{Ham2004} present an interesting technique to visualize small-world graphs using interactive clustering and an enhanced force-directed algorithm \cite{Eades00navigatingclustered}. Auber {\it et al.} \cite{Auber2003} present a work on the same theme using the clustering index metric \cite{Watts2003}. For the problem of non-clustered drawing, Harel and Yehuda \cite{Harel2002} describe an efficient method based on the embedding of graphs in high-dimensional spaces followed by a PCA (Principal Component Analysis) dimensionality reduction to two or three dimensions.}

Huang and Nguyen \cite{Huang2007} present a methodology for visualizing hierarchical graphs. They introduce an efficient layout scheme, being able to scale to tens of thousands of nodes. Different from our work, they do not integrate the relationships lost after the hierarchy generation; neither do they use a proper data structure, so their system is limited to main memory. Papadopoulos and Voglis \cite{Papadopoulos2005} propose a drawing method based on graph modular decomposition \cite{Dahlhaus2001}. Their work does not present a complete system, but a description of how to arrange the modules of a graph according to hierarchical levels. In the GrouseFlocks system, Archambault {\it et al.}\cite{4447668} define metanodes and metaedges to introduce the same visualization paradigm that we employ in our proof of concept experiments; differently they focus on layout and interaction with one order of magnitude higher processing demands for smaller graphs. Generally, former works -- as those presented by Finocchi \cite{Finocchi2002} -- have not considered the issue of efficiently managing graph hierarchies, instead, they rely on {\it ad hoc} linear or matrix adjacency structures. The use of such structures leads to hierarchies that do not provide comprehensive graph relationship information, mostly due to the scalability shortcomings of these approaches. In the literature, the goal of authors has been aesthetics; while here, we aim at a model that is more suitable for large scale computation and mining.

In the specific field of {\it hierarchical graph navigation}, Buchsbaum and Westbrook \cite{338609} formally present the problem and provide a solution in which the graph hierarchy has one unique associated state that changes according to two possible transitions: {\it expand} and {\it contract}. In their model, the graph nodes and the nodes of the hierarchy are a single concept at different levels of abstraction. In another work, Raitner \cite{1535414}, along with an extensive research compilation, deals with the issue of dynamically editing the nodes that are under a sub-tree of the hierarchy structure. These two works are references for what is known as {\it graph view maintenance problem}. Differently to the {\it view maintenance} approach, we describe a framework that aims not only at hierarchical navigation, but at large graph processing by means of a data structure that can fully represent a graph by abstracting the fact that it is hierarchically partitioned. Our structure is based on three integrated concepts: graph hierarchy, subgraphs, and graph nodes; it can restore the adjacency information of a single graph node or compute the relationship of arbitrary graph partitionings with a fraction of the original graph in memory, defining a complete graph representation over a hierarchical structure.

In speaking about visual design, the field of tree-like visualization is long term now and has a great number of branches as compiled by Hans-J\"org Schulz at \url{http://treevis.net/}. In this scenario, the aim of our work is to propose processing techniques that fit to any tree-like design in the task of scalable hierarchical graph visualization. As so, GMine's visual appeal was conceived as a proof of concept of our intent, accordingly, it does not compete with more elaborated designs.\\

\noindent{\bf Graph Representation}\\
\noindent{Two classic data structures usually are used for graph representation: adjacency matrices and adjacency lists. Another possibility is to use Binary Decision Diagrams \cite{Akers1978}, which represent the nodes of the graph using binary sequences. This approach supports massive processing using less memory, however, the nodes can no longer be processed individually \cite{Gentilini2003}. These three techniques are limited to main memory, this is because they are plain and do not provide the benefits of optimized disk access offered by hierarchical structures. Another line of research considers out-of-memory algorithms \cite{Vitter2001}, according to which the graph is preprocessed for specific computations. Such algorithms minimize disk accesses, however the computation is not versatile and does not favor interaction. Finally, Davi \cite{Dalvi2008} define a representation for hierarchically partitioned graphs similar to our approach -- using the concepts of SuperNodes and SuperEdges; however, their representation is intended for completely different purposes -- keyword search over graphs.}\\

\noindent{\bf Graph Summarization}\\
\noindent{Besides the capability of globally analyzing large graphs, our system is complemented with the possibility of locally analyzing a subgraph that is part of a larger graph hierarchy. For this aim, we use a graph summarization method named Center-Piece Subgraph -- {\CEPS} \cite{Tong06KDD}  -- adapted for visual interaction and presentation, and embedded at the leaves of our graph representation. A center-piece subgraph contains the collection of paths connecting a subset of nodes of interest. It has been shown that the center-piece subgraph can discover a collection of paths rather than a single path, and is preferable to other methods on describing the multi-faceted relationship between entities in a social network. The {\CEPS} method uses random walk with restart to calculate an importance score between graph nodes. Random walks refer to stochastic processes where the position of an entity, in a given time, depends on its position at some previous time. There are many applications using random walk methods, including PageRank \cite{54}, cross-modal multimedia correlation discovery \cite{MMG}, and neighborhood formation in bipartite graphs \cite{56}.}

The MING approach \cite{Kasneci2009} extends {\CEPS}' ideas to disk-resident graphs and to the Entity-Relationship database context providing the {\it IRank} measure to capture the informativeness of related nodes. In recent works, Patel {\it et al.} conducts a research effort on how to produce graph summaries. Their SNAP summarization uses node attributes combined to the implicit domain knowledge embedded in the graph structure and content \cite{Tian2008}; further in this line \cite{Zhang2010}, an automatic numerical categorization produces multiple summaries compared by means of a measure of interestingness.

{\CEPS} also relates to the concept of {\em ``goodness'' of a connection subgraph}. The two most natural measures for goodness are the shortest distance and the maximum flow. However, as pointed out by Faloutsos {\it et al.} \cite{53}, both measurements fail to capture some preferred characteristics for social networks. A more related closeness (distance) function is proposed by Palmer and Faloutsos \cite{Palmer03Electricity}. However, it cannot describe the multi-faceted relationship that is essential in social networks. In \cite{53}, Faloutsos {\it et al} propose a method based on electricity current, in which the graph is seen as an electric network. By applying +1 voltage to one query node and setting the other query nodes at 0 voltage, their method chooses the subgraph which delivers maximum current between the query nodes. The delivered current criterion can only deal with pair wise source queries, which is a special case of the {\CEPS} graph summarization.

\section{SuperGraphs and the Graph-Tree}
\label{sec:Terminology}

Our first contribution is an original formalization of graph hierarchies engineered to support processing and presentation. We define the SuperGraphs concept, an abstraction that converges to an implementation model we have named Graph-Tree. While SuperGraphs formalize the essentials of the Graph-Tree, the Graph-Tree incorporates the SuperGraph abstraction. SuperGraphs extend previously-proposed graph hierarchy representations -- Section \ref{sec:Connectivity} -- while the Graph-Tree instantiates it in a way that is propitious for efficient computation -- Section \ref{sec:Performance} and interactive presentation -- Section \ref{sec:SuperGraph_Visualization_in_GMine}. 

The closest work to the ideas of SuperGraph and Graph-Tree was proposed by Abello {\it et al.} \cite{Abello2006}. Their work formalizes a hierarchy tree, whose data structure is based on what they name antichains -- sets of nodes such that no two nodes are ancestors of one another. Their formalization parallels with ours by the concept of {\it macro} -- similar to the terminology {\it super}, used along this work. Their structure stores a static set of macro (super) edges between the macro (super) nodes of the hierarchy; differently, our data structure introduce the {\it Connectivity} computation, a dynamic means to determine macro (super) edges between arbitrary macro (super) nodes, even for the leaves (solely nodes). The originality of our approach is that the graph hierarchy is not available only for visual interaction; it can be used for processing at any level of the tree just as if the original graph was a thorough plain representation. This is possible due to the connectivity computation embedded in the Graph-Tree, as defined in section \ref{sec:Connectivity}.

\subsection{Graph-Tree Structure Formalization}
\label{sec:formalization}

\noindent{For the purpose of formalizing the Graph-Tree structure \footnote[1]{For a standard formalism on clustered graphs, see the seminal work of Harel \cite{Harel87}.}, following we define a set of abstractions that encompass its engineering, starting by the notion of SuperGraph. The underlying data beneath a SuperGraph is a graph $G=\{V,E\}$ -- with $|V|$ nodes and $|E|$ edges -- but a SuperGraph presents a different abstract structure. It is based on the observation that the entities in a graph can be grouped according to the relationships that they define. This concept allows us to work with a graph as a set of partitions hierarchically defined. In the following, we define the constituents of a SuperGraph, illustrating them with the example in Figure \ref{fig:fig2}.}

\begin{figure}[htb]
        \centering         
\includegraphics[width=0.4\textwidth]{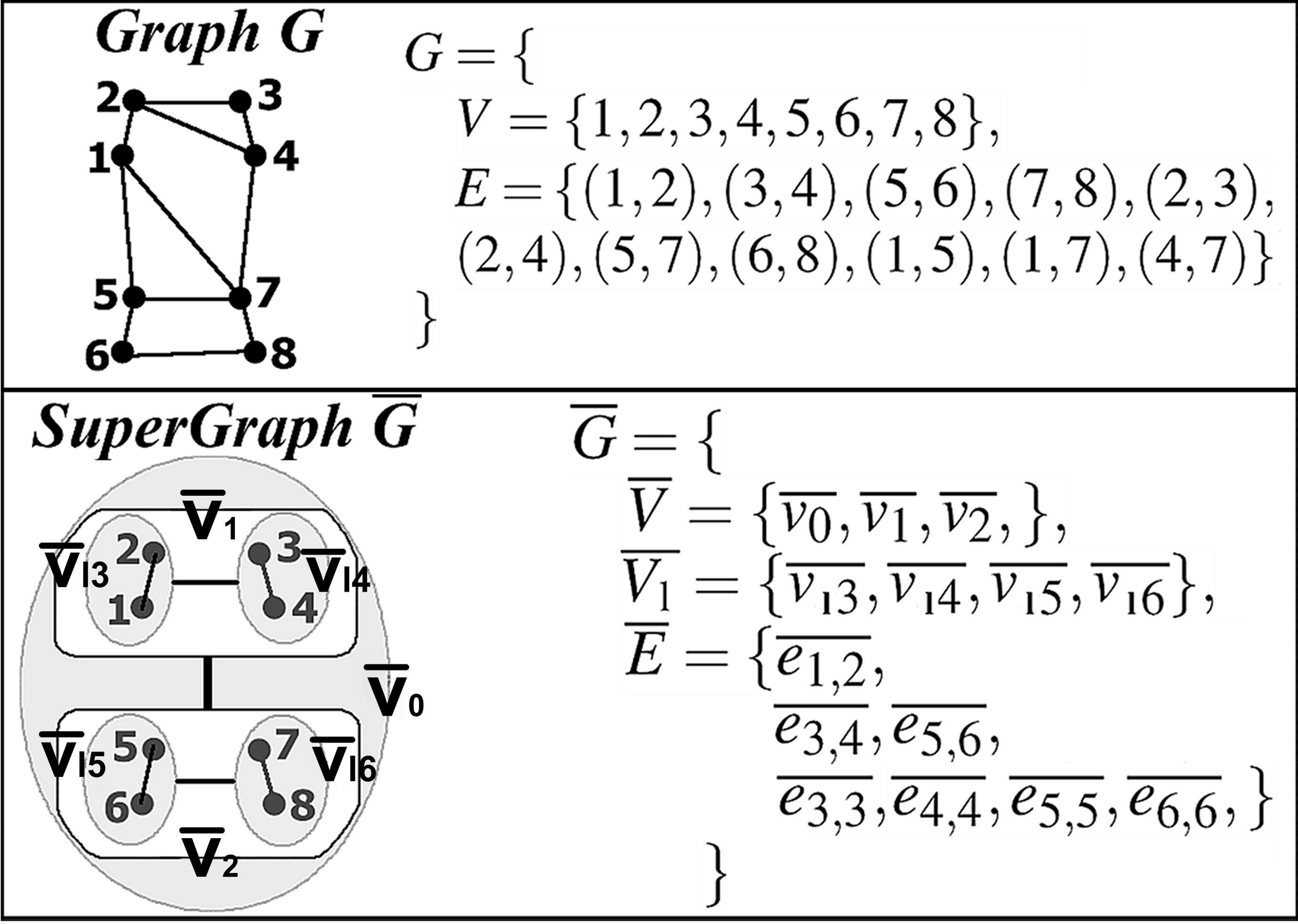}
        \caption{Example of a Graph and the respective SuperGraph. For the SuperGraph $\overline{G}$, $\overline{V}$ is the set of SuperNodes, $\overline{V_l}$ is the set of LeafSuperNodes, and $\overline{E}$ is the os SuperEdges.}
        \label{fig:fig2}
\end{figure}

\addtocounter{definitionsCounter}{1}
\noindent{{\bf Definition \arabic{definitionsCounter}: [SuperGraph]}\
Given a finite undirected graph $G=\{V,E\}$, with no loops nor parallel edges, a SuperGraph is defined as
$\overline{G}=\{\overline{V},\overline{V_l},\overline{E}\}$, where $\overline{V}$ is a set of {\SuperNode}s $\overline{v}$, $\overline{V_l}$ is a set of {\CommunitySuperNode}s $\overline{v_l}$, and $\overline{E}$ is a set of SuperEdges $\overline{e}$. In the following, we define {\CommunitySuperNode}, {\SuperNode}, and SuperEdge.}

\addtocounter{definitionsCounter}{1}
\noindent{{\bf Definition \arabic{definitionsCounter}: [{\CommunitySuperNode}]}\
Given a subset of graph nodes $V' \subset V$, a {\it \CommunitySuperNode} $\overline{v_l}$ is defined as the subgraph $G'=\{V',E'\}$, where $E'=\{(u,v)|(u,v)\in E\ and\ u, v \in V'\}$.}

\addtocounter{definitionsCounter}{1}
\noindent{{\bf Definition \arabic{definitionsCounter}: [\SuperNode]}\
A {\it {\SuperNode}} $\overline{v}$ is recursively defined as a set $\overline{V'}$ of {\SuperNode}s, or {\CommunitySuperNode}s, $\overline{v_i}$, plus a set $\overline{E'}$ of SuperEdges $\overline{e_{ij}}$. As follows:}

\begin{equation}
\begin{array}{c}
\overline{v}=\{\overline{V'}=\{\overline{v_0},\overline{v_1},...,\overline{v_{(|\overline{V'}|-1)}}\},\\ \ \ \ \ \ \ \ \ \ \ \ \ \ \ \overline{E'}=\{\overline{e_{ij}}=(\overline{v_i},\overline{v_j})|
\overline{v_i},\overline{v_j} \subset \overline{V'}\}\}
\end{array}
\label{eq:SuperNode}
\end{equation}

\noindent{where $\overline{v_i}$ can be either a {\SuperNode} or a {\CommunitySuperNode}; the concept of \emph{SuperEdge}, $\overline{e}$, is introduced later in the next subsection. Figure \ref{fig:fig2} illustrates the concepts of \emph{\SuperNode} and {\CommunitySuperNode}.}

Note that SuperNode and LeafSuperNode correspond to ``nodes'' in the hierarchy defined in a Graph-Tree. They are not to be confused with the individual {\it graph nodes} of the underlying graph.

\subsection{Basic definitions of the SuperGraph}
\label{sec:operations}
\noindent{The SuperGraph abstraction naturally lends to a novel tree-like model that we call Graph-Tree. Following, we present the basic operations for the Graph-Tree to work.}\\

\addtocounter{definitionsCounter}{1}
\noindent{{\bf Definition \arabic{definitionsCounter}: [Coverage of a {\SuperNode}]}\
Given a {\SuperNode} $\overline{v}=\{\overline{V'},\overline{E'}\}$, the coverage of $\overline{v}$ is given by the recursive definition:}

\begin{eqnarray}
Coverage(\overline{v})=
\begin{cases}
V',\ if \ \overline{v}\ is \ a \ \CommunitySuperNode \cr
\bigcup Coverage(\overline{v_i}),\ otherwise \cr
\end{cases}
\label{eq:Coverage}
\end{eqnarray}

\noindent{where $\overline{v_i} \ \in \ \overline{V'},\ 0 \leq i \leq |V'|-1$.}

The coverage of a {\SuperNode} corresponds to the graph nodes that comprehend its community. At the leaves, a community is a subgraph and, at the root, the community is the entire graph.

\addtocounter{definitionsCounter}{1}
\noindent{{\bf Definition \arabic{definitionsCounter}: [Parent(s) of a {\SuperNode}]}\
We refer to the parent of a {\SuperNode} $\overline{w}$ as $Parent(\overline{w})=\overline{v}=\{\overline{V'},\overline{E'}\}$ if $\overline{w} \in \overline{V'}$. We refer to the set of ancestors of a {\SuperNode} $\overline{w}$ as the set $Ancestors(\overline{w})=\{\overline{v}|\overline{v} \in \overline{V}\ and\ \overline{w} \in coverage(\overline{v})\}$. Similarly, two {\SuperNode}s (or
{\CommunitySuperNode}s) are \emph{siblings} if they have the same parent {\SuperNode}.}

\addtocounter{definitionsCounter}{1}
\noindent{{\bf Definition \arabic{definitionsCounter}: [SuperEdges]}\
A SuperEdge represents all
the edges $(u,v) \in E$ that connect graph nodes from a {\SuperNode}
$\overline{v_i}$ to graph nodes from {\SuperNode} $\overline{v_j}$.
A SuperEdge $\overline{e_{kk}}$ for a {\CommunitySuperNode}
$v_{lk}=\{V_k',E_k'\}$ holds all the edges that interconnect graph nodes in the {\CommunitySuperNode} $v_{lk}$, that is, all the edges in
$E_k'$.  Formally, the SuperEdge between {\SuperNode}s
$\overline{v_i}$ and $\overline{v_j}$ is defined as follows:}
\begin{equation}
\begin{array}{c}
   SuperEdge(\overline{v_i}, \overline{v_j})=\overline{e_{ij}}=\{e=(u,v)| (u,v) \in E, \cr
   \ u \in Coverage(v_i)\ and\ v \in Coverage(v_j)\} \cr
\end{array}
\label{eq:SuperEdge}
\end{equation}

\addtocounter{definitionsCounter}{1}
\noindent{{\bf Definition \arabic{definitionsCounter}: [Weight of a SuperEdge]}\
The weight of a SuperEdge is the sum of the weights of its edges.}

\addtocounter{definitionsCounter}{1}
\noindent{{\bf Definition \arabic{definitionsCounter}: [Internal Edge]}\ Given a {\SuperNode} (or a {\CommunitySuperNode}) $\overline{v}$, an edge $e$ is called an {\it internal edge} of $\overline{v}$ if $source(e) \in Coverage(\overline{v})$ and $target(e) \in Coverage(\overline{v})$. The internal edge $e$ can be resolved within the coverage of $\overline{v}$. For simplification, given an edge $(u,v)$, $u=source(e)$ and $v=target(e)$, even if the edges are undirected.}

\addtocounter{definitionsCounter}{1}
\noindent{{\bf Definition \arabic{definitionsCounter}: [External Edge]}\ An edge $e$ is called an {\it external edge} of $\overline{v}$ if $source(e) \in Coverage(\overline{v})$ and $target(e) \not \in Coverage(\overline{v})$. The external edge $e$ cannot be resolved within the Coverage of $\overline{v}$.}

\addtocounter{definitionsCounter}{1}
\noindent{{\bf Definition \arabic{definitionsCounter}: [Open Node]}\ A graph node $v \in Coverage(\overline{v})$ is called an {\it open node} of $\overline{v}$ if there exists an external edge $e$ in the set of {\it external edges} of $(\overline{v})$ where $source(e)=v$. We denote the set of all the {\it open nodes} of a {\SuperNode} $\overline{v}$ as $OpenNodes(\overline{v})$.}\\

With these basic definitions in mind, the engineering of the Graph-Tree can be better understood by tracing its process of construction, as presented in the next section.

\subsection{Construction of the GraphTree}
\label{sec:GraphTree}

\noindent{In this section we describe how to build a Graph-Tree. We illustrate the process in order to clarify its structure and the information it manages.}\\

\noindent{\textbf{Hierarchy construction}}\\
The choice for a specific graph partitioning is independent of the Graph-Tree methodology. The partitioning can be part of a dataset with a hierarchical structure, or it can be achieved via automatic partitioning. For automatic partitioning, in GMine, we recursively apply the k-way graph partitioning known as METIS, as described by Karypis and Kumar \cite{1}. We perform a sequence of recursive partitionings. Each recursion generates $k$ partitions to form the next level of the tree, a process that repeats until we get the desired number of $h$ hierarchy levels. For each new set of partitions (subgraphs), new subtrees are embedded in the Graph-Tree. At the end of the process, references to the subgraphs are kept at the leaves. From the storage point of view, the tree-structure is kept on main memory, while the subgraphs are kept on disk, being read only when necessary.\\

\begin{figure}[htb]
        \centering         
\includegraphics[width=0.49\textwidth]{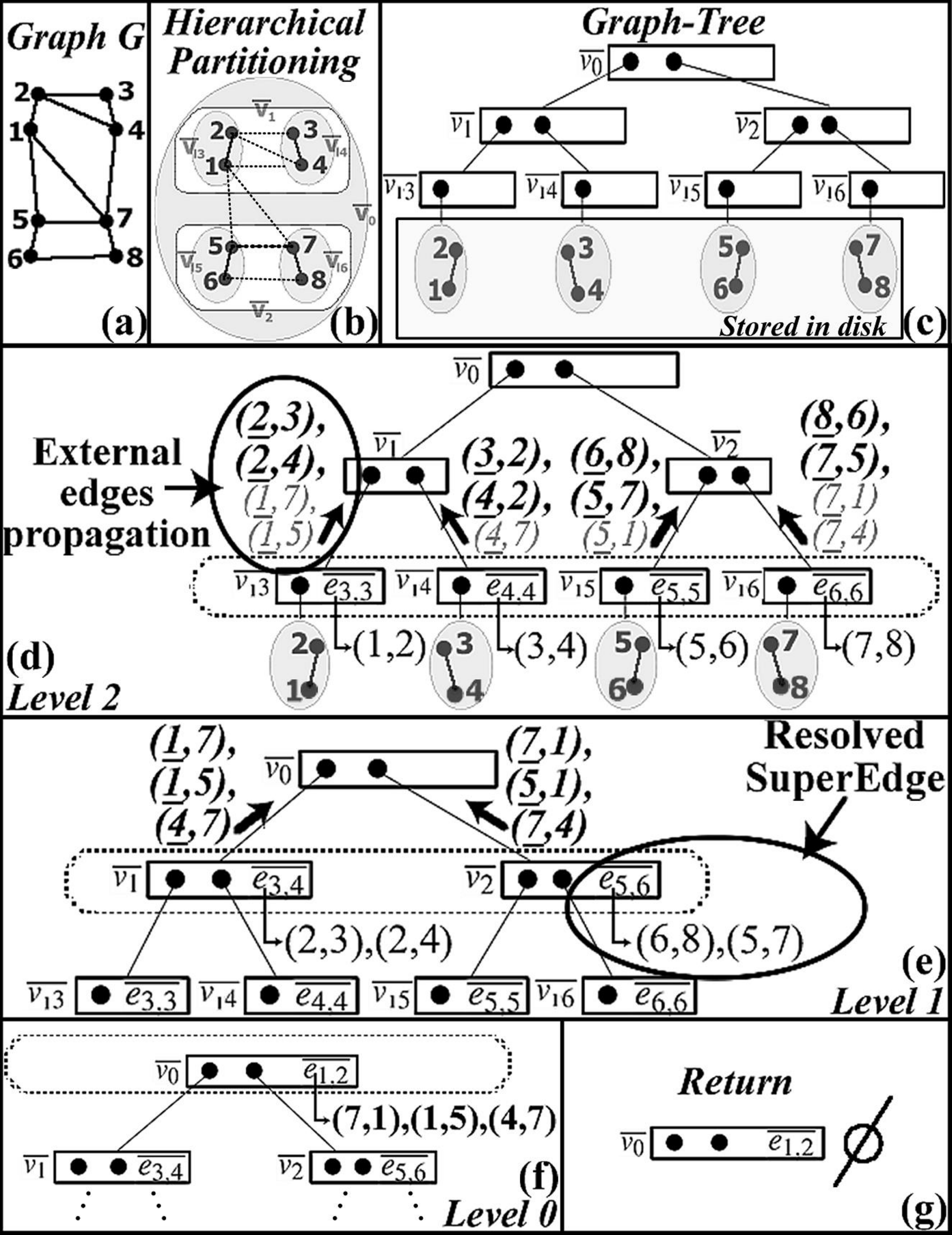}
        \caption{Filling a Graph-Tree. From (a) to (c), hierarchical partitioning and empty Graph-Tree creation. From (d) to (g), illustration of the FillGraphTree algorithm (Algorithm \ref{alg:TreeFilling}).}
        \label{fig:fig4}
\end{figure}

\noindent{\textbf{Filling the Graph-Tree {\SuperNode}s}}\\
After obtaining a hierarchy, it is necessary to fill the {\SuperNode}s of the tree with their SuperEdge and {\it open nodes} information. In Algorithm \ref{alg:TreeFilling}, the Graph-Tree is recursively traversed bottom-up along its levels. Initially the LeafSuperNodes are filled with references to the subgraphs stored on the disk. Then, the algorithm proceeds to upper levels, where the {\it external edges} propagated from lower levels are used to resolve the SuperEdges and to track the {\it open nodes}.

Figure \ref{fig:fig4} illustrates this process. We start with graph $G$, which is partitioned to create the Graph-Tree with empty {\SuperNode}s (see Figures \ref{fig:fig4}(a), \ref{fig:fig4}(b) and \ref{fig:fig4}(c)). The bottom-up recursive process starts at the leaves, illustrated in Figure \ref{fig:fig4}(d). For this example, and for Figure \ref{fig:fig4}(e), boldface indicates matches between {\it external edges}, while gray edges indicate unresolved {\it external edges}. Underlined graph node id's indicate {\it open nodes} and the diagonal arrows depict the {\it external edges} propagated up the tree. Still in Figure \ref{fig:fig4}(d), it is possible to see the information propagated from {\SuperNode}s $\overline{v_{l3}}$ and $\overline{v_{l4}}$, which will be used in line 8 of Algorithm \ref{alg:TreeFilling} to find matches between unresolved external edges. Figure \ref{fig:fig4}(e) illustrates the crossing of the propagated data results in matches $(2,3)-(3,2)$ and $(2,4)-(4,2)$, stored in SuperEdge $\overline{e_{3,4}}$. Figure \ref{fig:fig4}(e) also shows the first SuperEdges among siblings, ($\overline{e_{3,4}}$ and $\overline{e_{5,6}}$). Figure \ref{fig:fig4}(f) shows the last SuperEdge storing the last set of edges between siblings. Figure \ref{fig:fig4}(g) shows the end of the process, when all the edges are spread along the data structure.

\begin{algorithm}[htb]
\SetLine
\KwIn{$Ptr$: pointer to the root of the Graph-Tree}
 FillGraphTree(Ptr)
 \Begin{
   \eIf{$Ptr$ is leaf}{
     Set the variable $Ptr \rightarrow filePath$ to the file of the corresponding subgraph;
   }{ 
     \For{each child $s_i$ of $Ptr$}{
       $FillGraphTree(s_i)$\;
       /*Recursively down the hierarchy*/
     }
     Instantiate a SuperEdge for each pair of children\;
     Find matches between the unresolved external edges from each pair of children\;
     Store matching edges in the SuperEdges\;
   }
   Use external edges to determine $Ptr$'s open nodes\;
   Propagate (unresolved) external edges to the parent\;
 }
 \caption{\emph{Algorithm to fill a Graph-Tree.}}
 \label{alg:TreeFilling}
\end{algorithm}

\subsection{SuperGraph Connectivity Computations}
\label{sec:Connectivity}

\noindent{In this section, our aim is to answer the questions raised in Section \ref{sec:Introduction} by dynamically restoring the original graph information.}

\subsubsection{SuperNodes Connectivity}
\label{subsec:sNodesConnectivity}

\noindent{The {\it connectivity} between two {\SuperNode}s in a hierarchy is the set of edges between them. For sibling {\SuperNode}s, their connectivity corresponds to the SuperEdge that interconnect them, readily available as part of the SuperGraph. For {\SuperNode}s that are not siblings, their connectivity must be traced.}

\addtocounter{definitionsCounter}{1}
\noindent{{\bf Definition \arabic{definitionsCounter}: [{\SuperNode}s Connectivity]}  Given a SuperGraph $\overline{G}=\{\overline{V},\overline{V_l},\overline{E}\}$ and two {\SuperNode}s $\overline{v_i}$ and $\overline{v_j} \in \overline{G}$, the {\em SuperNodes Connectivity} for the pair $(\overline{v_i}$, $\overline{v_j})$ is the set of edges $SNC(\overline{v_i}$, $\overline{v_j})=\{e|source(e) \in Coverage(\overline{v_i})\ and \ target(e) \in Coverage(\overline{v_j})\}$.
  
The challenge is how to trace the connectivity between arbitrary {\SuperNode}s without having to cross the SuperGraph with the graph that originated it. To do so, we benefit from the SuperGraph definitions of the former subsection in order to calculate the connectivity between {\SuperNode}s.

\noindent{{\bf Proposition 1: [All possible connecting edges]}\  Given any two {\SuperNode}s $\overline{v_i}$ and $\overline{v_j}$, the complete set of all possible edges connecting $\overline{v_i}$ to $\overline{v_j}$ is given by the Cartesian product $OpenNodes(\overline{v_i})\ \times\ OpenNodes(\overline{v_j})$.}

\noindent{{\bf Proposition 2: [Connecting edges from the common parent]}\ The set of edges that connect any two {\SuperNode}s $\overline{v_i}$ and $\overline{v_j}$ is a {\em subset} of the unique SuperEdge $\overline{e_{gh}}$ connecting {\SuperNode}s $\overline{v_g}$ and $\overline{v_h}$, where $\overline{v_g} \in Ancestors(\overline{v_i})$ and $\overline{v_h} \in Ancestors(\overline{v_j})$, so that $\overline{v_f}=Parent(\overline{v_g})=Parent(\overline{v_h})$. Intuitively, $\overline{v_f}$ is the first common parent of $\overline{v_i}$ and $\overline{v_j}$; $\overline{v_g}$ and $\overline{v_h}$ are sibling {\SuperNode}s under $\overline{v_f}$ and are ``ancestors'' of $\overline{v_i}$ and $\overline{v_j}$, respectively.\\

From propositions 1 and 2, it becomes possible to calculate the connectivity between two {\SuperNode}s based on set operations, as follows.\\

\noindent{{\bf Proposition 3: [Computing {\SuperNode}s Connectivity $SNC(\overline{v_i},\overline{v_j})$]}\  The set of edges $SNC(\overline{v_i},\overline{v_j})$ that connect any two {\SuperNode}s $\overline{v_i}$ and $\overline{v_j}$ is the intersection between the set of all possible edges between $\overline{v_i}$ and $\overline{v_j}$ (Proposition 1) and the superset that contains (but not only) the set of edges between $\overline{v_i}$ and $\overline{v_j}$ (proposition 2). Formally, the {\SuperNode}s Connectivity $SNC(\overline{v_i},\overline{v_j})$ is given by:

\begin{equation}
\label{eq:connectivity}
SNC(\overline {v_i } ,\overline {v_j } ) = \begin{array}{*{20}c}
   \begin{array}{l}
 \{ OpenNodes(\overline {v_i } )\ \times\ \\ 
 OpenNodes(\overline {v_j } )\}  \\ 
 \end{array}  \\
    \cap   \\
   \begin{array}{l}
 \{ \overline {e_{gh} } |\overline {v_i }  \in Coverage(\overline {v_g }), \\ 
 \overline {v_j }  \in Coverage(\overline {v_h } )\}  \\ 
 \end{array}  \\
\end{array}
\end{equation}

To see why proposition 3 is the case, we note that $\overline{e_{gh}}$=SuperEdge($\overline{v_g}$,$\overline{v_h}$) contains all the edges between $Coverage(\overline{v_g})$ and $Coverage(\overline{v_h})$, and therefore it is a superset of $SNC(\overline{v_i},\overline{v_j})$.

\subsubsection{Graph Nodes Connectivity}
\label{subsec:gNodesConnectivity}

\noindent{A graph hierarchy stores groups (partitions) of nodes that are interrelated. However, the relationships between graph nodes at different groups are not stored; we lose information when we alter the graph representation. In a SuperGraph, it is possible to determine the relationships relative to any graph node, which we define as follows:}

\addtocounter{definitionsCounter}{1}
\noindent{{\bf Definition \arabic{definitionsCounter}: [Graph Nodes Connectivity]} Given a SuperGraph $\overline{G}=\{\overline{V},\overline{V_l},\overline{E}\}$, a {\SuperNode} $\overline{v_i} \in \overline{G}$, and a graph node $v \in Coverage(\overline{v_i})$, the {\em Graph Nodes Connectivity} for $v$ (denoted as $GNC(v)$), is defined as the set of edges $e \in E$ connecting $v$ to all the other graph nodes that do not pertain to $\overline{v_i}$. That is, $GNC(v)=\{e|e \in E, source(e) = v\ and \ target(e) \in \{V-Coverage(\overline{v_i})\}\}$.

\noindent{{\bf Proposition 4:}\  If a graph node $v$ is an open node for a {\SuperNode} $\overline{v}$, then the set of ancestors $Ancestors(\overline{v})$ have all the SuperEdges that hold edges connected to $v$. Proposition 4 is a direct result from Definition 6.}

Following Proposition 4, if we know the set of ancestors and the set of open nodes of a {\SuperNode}, we can determine the relationships (external edges) of any graph node $v \in OpenNodes(\overline{v})$. A reference to the immediate parent at each {\SuperNode} is enough to define a recursive procedure to trace the external edges of any graph node $v$. Such procedure checks each parent {\SuperNode}, starting from the first parent above the leaves, up to the root. While $v$ is in the set of open nodes of the parent {\SuperNode} being checked, then there are still {\it external edges} to be traced.

In this section, we have presented the SuperGraph/Graph-Tree formalism, which carries an engineering that elegantly allows the construction of a graph hierarchy. It also predicts computation that can restore all the relationships of the original graph, and that can calculate relationships between {\SuperNode}s at any levels of the hierarchy. In section \ref{sec:Performance}, we demonstrate that the Graph-Tree can perform its computations with sub linear complexity, scaling to graphs that are really big.

\section{{\CEPS}: Center-Piece Subgraph}
\label{sec:CePs} 

\noindent{Although graph hierarchies can lessen the problem of globally inspecting large graphs, we have found that it is common to reach the bottom of the Graph-Tree and have a subgraph that presents more information than what is desired, in a layout that suffers with node overlapping. In this situation, although the user is able to compute, draw and interact with the graph nodes of a {\CommunitySuperNode}, there might still be too many edges and nodes, preventing examination. This happens naturally, either on large graphs or on moderate to small graphs.}

To remedy this problem, we benefit from the concept of {\it Center-Piece Subgraph} ({\CEPS} for short) to complement the analytical environment of GMine. A center-piece subgraph contains the collection of paths connecting a subset of graph nodes of interest. Using the {\CEPS} method, a user can specify a set of query graph nodes and GMine will summarize and present their internal relationship through a small (say, with tens of nodes), yet representative {\it connection subgraph}.

{\CEPS} aids on interaction by significantly reducing the number of edges and of nodes to be inspected; we can estimate its benefits analytically. For a complete graph $G'$ -- a worst case situation -- Figure \ref{fig:fig11}(a), one must manually check $|N|(N-1)/2$ edges in order to manually generate a center-piece subgraph, considering the edges node by node -- Figure \ref{fig:fig11}(b); while with {\CEPS}, only the nodes must be considered, and no edges at all. In respect to the number of nodes to be considered, with {\CEPS} this number decreases linearly with the number of nodes in the budget; for $b = 1$, the problem is similar to the manual inspection of the graph, which demands the consideration of all the $N$ nodes in $G'$. For $b=N-|Q|$, the problem requires the inspection of only $|Q|$ nodes -- possibly with $|Q| << N$; that is, one must only determine the source nodes that feed the algorithm -- Figure \ref{fig:fig11}(c), proceeding interactively to the user's demand -- Figure \ref{fig:fig11}(d). In other words, GMine brings interaction to the broadly studied problem of graph summarization, combining it to hierarchical graph visualization.

\begin{figure}[htb]
        \centering         
\includegraphics[width=0.49\textwidth]{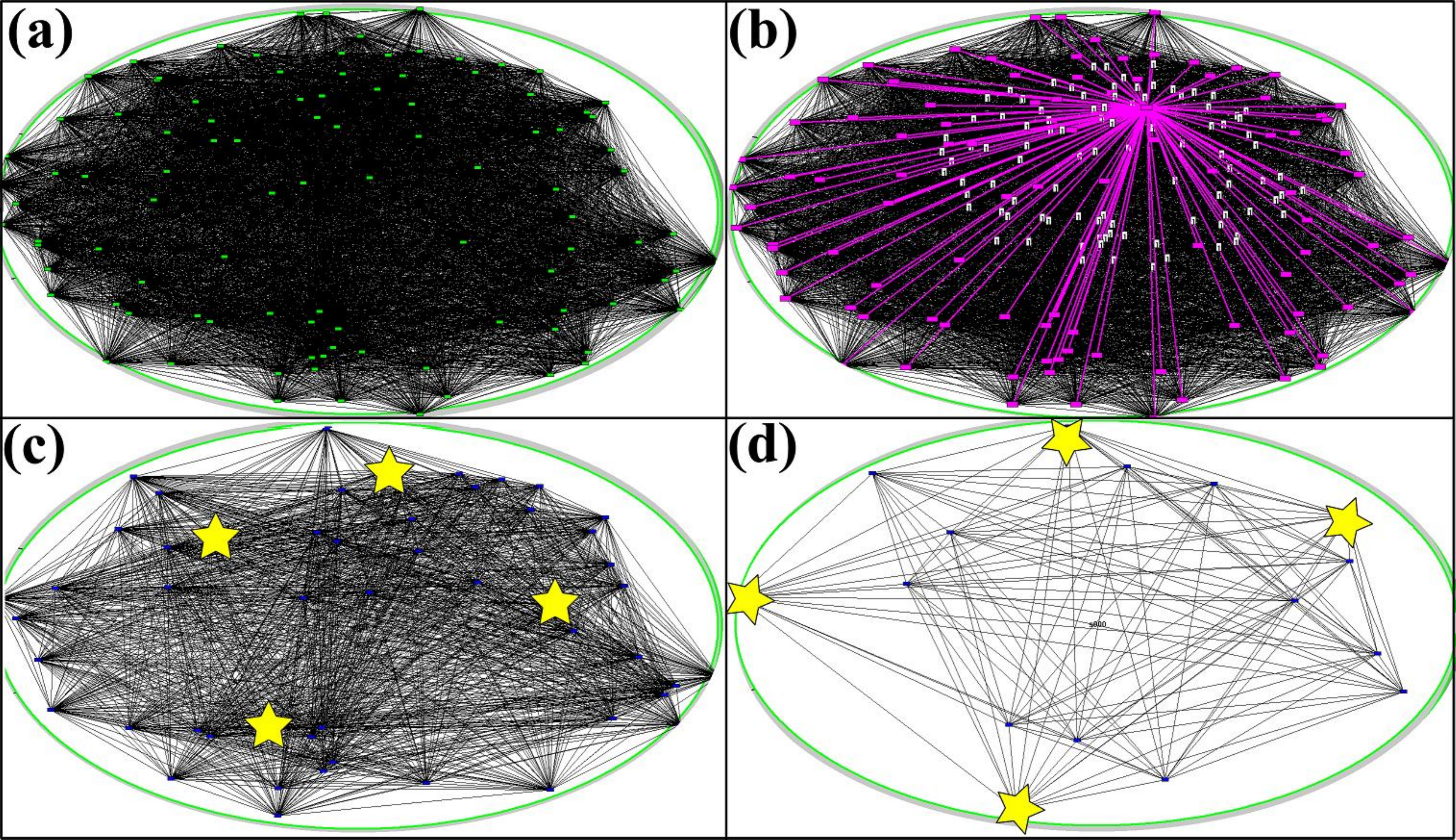}
        \caption{{\CEPS} visual summarization. (a) A complete graph problem -- 100 nodes and 4950 edges. (b) Inspection of the edges of a single node. (c) First summarization with $Q=4$ source query nodes and a budget of $b=50$ nodes. (d) Further summarization with $Q=4$ and $b=16$.}
        \label{fig:fig11}
\end{figure}

\subsection{{\CEPS} Overview}
\label{subsec:ceps_overview}

Given $Q$ graph nodes on a graph, how do we summarize the connectivity relationship among these nodes? The {\CEPS} technique proposes to represent such relationship with a {\em connection subgraph}. Such subgraph corresponds to the graph nodes that are center-piece and have direct or indirect connections to all, or most, of the nodes of interest. Formally, given $Q$ query nodes in a graph $\matW$=\{$V'$,$E'$\} ($\matW$ as a subgraph in a Graph-Tree), find the subset of nodes $CP \in V'$ that will determine an induced subgraph $CP \subset G'$ with budget $b$ (maximum $CP$ size in number of nodes) having strong connections to all $Q$ query nodes.

Following, we will use the symbology presented in Table~\ref{tab:cepsdefiniton}.

\begin{table}[htb]
 \caption{Symbols.}
 \begin{tabular}{|p{0.07\textwidth}|p{0.36\textwidth}|}\hline
  {\bf Symbol} & {\bf Description}\\\hline \hline
   $\matW$ & the subgraph of a given {\CommunitySuperNode}\\ \hline
         $N$ & total number of nodes in graph $\matW$\\ \hline
   $\QN$   & number of source query graph nodes \\ \hline
   $\QMySet = \{q_i\}$ & set of query graph nodes ($i=1,...,\QN$)\\ \hline
         $\unitvi$ & $N$-by-$1$ unit query vector all zeros except one at row $\qi$\\ \hline
         $\Hgraph$ & the induced {\CEPSfull} subgraph\\ \hline
 \end{tabular}
\label{tab:cepsdefiniton}
\end{table}

\noindent{A natural way to measure the validity of a subgraph $\Hgraph$ is to measure the goodness of the graph nodes it contains: the more ``good''/important nodes (with respect to the source queries) it contains, the better $\Hgraph$ is. Let us first define the goodness score for nodes. For a given graph node $j$, we have two types of goodness score:}

\begin{itemize}
\item{Let $\cijpar$ be the goodness score of a given graph node $j$ with respect to the query graph node $\qi$;}
\item{Let $\cqj$ be the goodness score of a given graph node $j$ w.r.t. the query set $\QMySet$.}
\end{itemize}

It follows that the goodness criterion for a $\Hgraph$ can be defined as:

\begin{equation}
g(\Hgraph)  =  \sum_{j\ \in\ nodes(\Hgraph)}{\cqj}
    \label{eq:gooddef}
\end{equation}

For this definition, there are two problems to achieve the center-piece subgraph: 1) how to define a reasonable goodness score $\cqj$ for a given graph node $j$; 2) how to quickly find a connection subgraph maximizing $g(\Hgraph)$.

\subsection{Goodness Score Calculation}
\label{subsec:goodness}

\noindent{The concepts for goodness score calculation are:}

\begin{itemize}
 \item{Let $\cij$ be the {\ssp} that a particle will find itself at node $j$, when it does random walk with restart (RWR) from a query node $\qi$.}
 \item{Let $\cqjQ$ be the {\pmetric}, that is, the steady-state probability that ALL $\QN$ particles, doing RWR from the query nodes of $\QMySet$, will all find themselves at node $j$ in the steady state.}
\end{itemize}

First, we want to compute the goodness score $\cijpar$ of a single graph node $j$, for a single query node $\qi$. To do so, we use random walk with restart from query node $\qi$. Suppose a random particle starts from node $\qi$, the particle iteratively transmits to its neighborhood with a probability that is proportional to the edge weight between them. Also, at each step, it has a probability $1-c$ to return to node $\qi$. In this conception, $\cijpar$ is defined as the {\ssp} $\cij$ that the particle will finally be at node $\qi$:

\begin{eqnarray}
\cijpar \triangleq \cij \label{eq:iNonedefinition}
\end{eqnarray}

    Formally, if we put all the $\cij$ probabilities into matrix form $\matC = [\cij]$, then

    \begin{eqnarray}
    \matC^T = c \matC^T \ \matWnorm + (1-c) \mat{E}
    \label{eq:RWRiN1}
    \end{eqnarray}
    where $\mat{E} = [\unitvi],\ for\ i=1,...,\QN$ is a $N$-by-$\QN$ matrix,
    $c$ is the fly-out probability, and $\matWnorm$ is
    the (column-) normalized adjacency matrix for graph $\matW$. The problem of determining $\matC^T$ can be solved in many ways - we choose the iteration method, iterating equation \ref{eq:RWRiN1} until convergence.

Once $\matC^T$ is ready, we want to combine the individual scores together to measure the importance for each graph node $j$ w.r.t. the whole query set $\QMySet$. The most common query scenario might be {\lqu given $\QN$ query nodes, find the subgraph $\Hgraph$ whose nodes are important/good w.r.t. ALL query nodes.\rqu} In this case, $\cqj$ should be high if and only if there is a high probability that all particles will finally meet at node $j$. This probability is given by:

\begin{eqnarray}
\cqj \triangleq \cqjQ = \prod_{i = 1}^{\QN}{\cijpar}
\label{eq:iNall}
\end{eqnarray}

The goodness score $\cqj$ of a given graph node $j$ w.r.t. the query set $\QMySet$ is the first step in order to calculate the induced center-piece subgraph $CP$. The next step is the {\em ``EXTRACT''} algorithm.

\subsection{The {\lqu EXTRACT\rqu} Algorithm}

\noindent{The {\lqu EXTRACT\rqu} algorithm takes as input the graph $\matW$, the importance/goodness score $\cqj$ on all nodes, and the budget $b$, and produces as output a small, undirected graph $\Hgraph$. The basic idea is as follows: 1) instead of trying to find an optimal subgraph maximizing $g(\Hgraph)$ directly, we decompose it, finding key paths incrementally; 2) by sorting the graph nodes in order, we can quickly find the key paths by dynamic programming in the acyclic graph.}

Before presenting the algorithm, we require the following definitions:

\addtocounter{definitionsCounter}{1}
\noindent{{\bf Definition \arabic{definitionsCounter}:} A graph node $u$ is called {\em specified downhill} from node $v$ w.r.t. source $\qi$ ($v \to _{i} u$) if $r(i,v) > r(i,u)$.}

\addtocounter{definitionsCounter}{1}
\noindent{{\bf Definition \arabic{definitionsCounter}:} A {\em specified prefix path} $P(i,u)$ is any downhill path that starts from source $\qi$ and ends at node $u$; that is, $P(i,u) = (u_0,u_1,...,u_n)$ where $u_0
= \qi, u_n = u$, and $u_j \to _{i} u_{j+1}$, for every $j$.}

\addtocounter{definitionsCounter}{1}
\noindent{{\bf Definition \arabic{definitionsCounter}:}\ The {\em extracted goodness} is the total goodness score of the nodes within the subgraph $\Hgraph$: $CF(\Hgraph) = \sum_{j \in \Hgraph}\cqj$.}

\addtocounter{definitionsCounter}{1}
\noindent{{\bf Definition \arabic{definitionsCounter}:}\ We define an {\em extracted matrix} as the matrix whose $(i,u)$ element, $C_s(i,u)$, corresponds to the extracted goodness score from a source graph node $\qi$ to node $u$ along the prefix path $P(i,u)$ such that:}

\begin{enumerate}
\item $P(i,u)$ has exactly $s$ nodes not in the present output graph $\Hgraph$, and
\item $P(i,u)$ extracts the highest goodness score among all such paths that start from $\qi$ and end at $u$.
\end{enumerate}

In order to discover a new path between the source $\qi$ and a destination node $\pd$, we arrange the nodes in descending order of $\cijpar (j=1,...,n)$: \{$u_1 = \qi,u_2,u_3,...,pd = u_n$\}. Note that all nodes with smaller $\cijpar$ than $r(i,\pd)$ are
ignored. Then we fill the extracted matrix $C$ in topological order so that when we compute $C_s(t,u)$, we have already computed $C_s(t,v)$ for all $v \to _{i} u$. On the other hand, as the subgraph is growing, a new path may include nodes that are already in the output subgraph. Our algorithm will favor such paths. The complete algorithm to discover a single path from source node $\qi$ and the destination node $\pd$ is given in Algorithm \ref{alg:pathdis}. Based on the previous preparations, the {\extract} algorithm is given in Algorithm \ref{alg:genedis}.

\begin{algorithm}[htb]
\SetLine
 Let $\QMySet$ be the set of query nodes\;
 Let $len$ be the maximum allowable path length\;
 Let $\mathcal{S}$ be a set of nodes \{$u_1 = \qi,u_2,u_3,...,pd = u_n$\}, where  $u_{k} \to _{i} u_{k+1}$, for $k=1,\ldots,(n-1)$.
 
 \For{$j \leftarrow [1,...,n]$}{
   Let $v = u_j$\;
   \For{$s \leftarrow [2,...,len]$}{
     \eIf{v is already in the output subgraph}{
       $s' = s$\;
     }{ 
       $s' = s-1$
     }  
     Let $C_s(i,v) = max_{u|u \to _{i},v}{(C_{s'}{(i,u)}+r(\QMySet,v))}$
   }
 }
 \KwResult{The path maximizing $C_s(i,\pd)/s$, where $s \ne 0$}
 \caption{Single Key Path Discovery (from node $i$ to node $\pd$).}
 \label{alg:pathdis}  
\end{algorithm}

\begin{algorithm}[htb]
\SetLine
 Initialize output graph $\Hgraph$ as an empty graph\;
 Let $len$ be the maximum allowable path length\;
 \While{$\Hgraph$ is not big enough (i.e., within the budget $b$)}{
   Pick up destination node $pd$: $pd  =  argmax_{j \notin {\Hgraph}}{\cqj}$\;
   \For{each source node $\qi$}{
     Use Algorithm \ref{alg:pathdis} to discover a key path $P(\qi,pd)$\;
     Add $P(\qi,pd)$ to $\Hgraph$\;
     /*Duplicate path nodes are detected and merged when paths are added to CP*/
   } 
 } 
 \KwResult{The final $\Hgraph$}
 \caption{The {\extract} Algorithm.}
 \label{alg:genedis}
\end{algorithm}

The EXTRACT algorithm joins all the formalism presented in this section, the goal is to systematically compute the Center-Piece Subgraph that best summarizes a graph of interest. In Section \ref{sec:ceps_accuracy} we present experiments attesting its accuracy and in Section \ref{sec:SuperGraph_Visualization_in_GMine} we demonstrate it.

\section{Graph Tree Performance}
\label{sec:Performance}

\noindent{Now, we present performance tests for calculating the SuperNode Connectivity (SNC) (Section \ref{subsec:sNodesConnectivity}) and the Graph Nodes Connectivity (GNC) (Section \ref{subsec:gNodesConnectivity}). We demonstrate that its performance surpasses that of classic adjacency lists and of relational databases.}\\

\noindent{\textbf{Complexity Analysis}}\\
\noindent{Considering a k-way partitioned Graph-Tree with $tn$ nodes (consisting of $sn$ {\SuperNode}s and $lsn$ LeafSuperNodes), the height of the tree is given by $h = \lceil log_k(tn(k-1)+1) \rceil$ -- root is level $1$; and the number of SuperEdges at level $l$ is given by $se(l,k) = (k!/(2!(k-2)!))$. In the configuration of a complete Graph-Tree, $sn = \displaystyle\sum_{i=1}^{i-1} k^{h-1}$ {\SuperNode}s; $lsn = k^{h-1}$ LeafSuperNodes; let $p=|V|/lsn$ be the number of graph nodes per subgraph, $d = |E|/|V|$ be the average degree of a graph node and $r$ be the expected ratio of external edges per graph node, $1/d \leq r < 1$ for $d > 1$. Also, let $f$ be the expected number of edges in a SuperEdge $\overline{e}$, where $Level(\overline{e})$ corresponds to the level of the SuperNodes that define $\overline{e}$; more especifically, $f(Level(\overline{e})) = \frac{|E|*r}{se(Level(\overline{e}),k)}$ if $Level(\overline{e}) = 1$ and $f(Level(\overline{e})) = \frac{f(Level(\overline{e})-1)*r}{se(Level(\overline{e}),k)}$ else.}

With these parameters, the complexity time for SuperNodes Connectivity, $SNC(\overline{v_i},\overline{v_j})$ is determined by the following factors: (1) time to search for the first common parent, $\overline{v_f}$, of $\overline{v_i}$ and $\overline{v_j}$, (2) time to search for the pair of siblings ($\overline{v_g},\overline{v_h}$) beneath $\overline{v_f}$ in the path to $\overline{v_i}$ and $\overline{v_j}$, (3) time to search for the SuperEdge($\overline{v_g}$,$\overline{v_h}$), and (4) time to perform the verification of which of the edges of SuperEdge($\overline{v_g}$,$\overline{v_h}$) pertain to the set of possible edges in between $\overline{v_i},\overline{v_j}$. The time complexity comes from $(3*h)+(k)+(2*f*r)$, where $k$ and $r$ are constants of the underlying graph, and $h$ is logarithmic; thus, the complexity is $O(f)$, where $f$, the expected number of edges in a SuperEdge, is a very small fraction of the number of edges $|E|$.

The Graph Nodes Connectivity, $GNC(v)$, is given by the time to trace the path from $v$ to the root; at each level up to the root, it takes the hash time to verify if $v$ is still an open node and, in each of the elements in the set of $k-1$ SuperEdges at a given level, it takes the hash time to track the edges that have $v$ as an endpoint. Thus, the time complexity comes from $(h)*(c)*(c*k)=h*c^2*k$; where $k$ is a constant, $c$ refers to the hash time assumed to be constant, and $h$ is logarithmic. Then, the chief term is $h$ and the complexity is logarithmic $O(h)$ for GNC.\\

\noindent{\textbf{Memory Consumption}}\\
\noindent{Since the Graph-Tree keeps leaf nodes on disk, it provides significant memory gains compared to the adjacency list. This gains depends on factor $r$, the expected ratio of external edges per graph node; the lower the value of $r$ the higher are the memory gains because more edges will be on disk and not on memory. In Figure \ref{fig:fig12} we present a comparative plot of the memory load for both the Graph-Tree and the adjacency list for a not favorable value of $r = 0.6$.}

\begin{figure}[htb]
\centering
\includegraphics[width=0.5\textwidth]{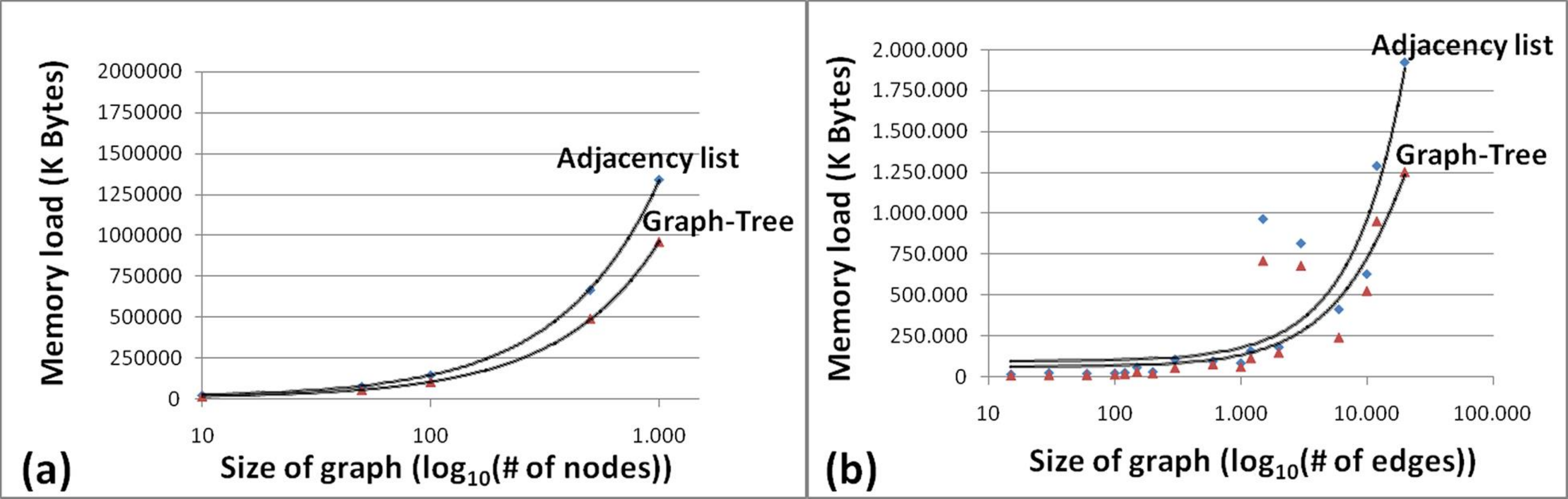}
\caption{Memory consumption. (a) Memory load in function of the number of nodes - log plot. (b) Memory load in function of the number of edges - log plot.}
\label{fig:fig12}
\end{figure}

\noindent{\bf Experiments Setting}\\
\noindent{We use synthetic graphs with varying number of nodes and average edge degree. We used graphs with 5K, 10K, 50K, 100K, 500K and 1M nodes with average edge degrees of 3, 12 and 20 edges per graph node; a total of 18 graphs whose number of edges ranges between 15K and 20M edges. We recursively break the graphs at up to $5$ levels and $5$ partitions per level, depending on the experiment, ranging from $2$ to $5^{5-1} = 3125$ partitions. We perform the experiments in a personal computer with a 3GHz processor, 4 MB L1 cache, 4 GB 500 MHz memory and a 5400 rpm 500 GB disk device. The entire experiment (data, code, software, performance measures and details) is available at \url{http://www.cs.cmu.edu/~junio}.

The goal is to observe the complexity cost using the wall-clock time necessary to calculate SNC and GNC. The SNC cost is chiefly determined by the expected number of edges ($f$) between the SuperNodes involved in the computation; so we vary this number from $500$ to $80K$ edges. The GNC cost is chiefly determined by the tree height ($h$) where a graph node lies; we use up to $5$ levels from trees that represent small to large scale graphs. We perform both all the above experiments for the Graph-Tree and the adjacency list and the first 12 of them with the DB2 commodity database.

The Graph-Tree was implemented following Section \ref{sec:Terminology} definitions so that besides a SuperGraph it also provides SNC and GNC functionalities. The adjacency list implementation was made on top of the {\it GraphGarden} graph library, under custody of researcher Jure Leskovec (\url{http://www.cs.cmu.edu/~jure/}). The graph nodes in the list are labeled according to the graph partitioning that they belong to. For maximum performance, the adjacency list uses hash mapping so that the retrieval of a given graph node is done in hash time. We also configured a relational database for the experiment. Its schema defines relations among graph nodes and SuperNodes allowing hierarchical management and SuperNodes' coverage computation. The database uses indexes for optimized searches and redundant information to reduce disk accesses.

\noindent{\textbf{Performance on SNC Computation}}\\
\noindent{The experiments confirmed the analytical expectations for the three different methodologies. The commodity database performance, despite its optimization, declines due to the nested SQL queries necessary for the SNC computation, what implies in random disk accesses. The database performance was one order of magnitude worse than the other two techniques. In turn, the adjacency list performance showed to be linear with the number of nodes and edges, reaching a reasonable performance at the cost of massive memory consumption. The Graph-Tree, on the other hand, is less sensible to these factors, having its performance determined by the size of the answer -- that is, the number of edges found in between two arbitrary SuperNodes, a fraction of the graph size (see the analytical calculus of $f$ in subsection {\it Complexity Analysis}).}

We note that the different natures of these two techniques ask for specific testing configurations. In Figure \ref{fig:fig8}(a), the parameters of interest are the number of nodes and edges; there we can verify how the adjacency list is more affected by the size of the graph than the Graph-Tree. In Figure \ref{fig:fig8}(b) the parameter of interest is $f$, calculated for several variations of the $18$ experimental graphs partitioned according to different levels and numbers of partitions per level. Along with Figure \ref{fig:fig8}(b), Figures \ref{fig:fig8}(c) and \ref{fig:fig8}(d) are intended to elucidate how the measures in Figure \ref{fig:fig8}(b) were performed; Figure \ref{fig:fig8}(c) shows that the number of graph nodes ranged from $5K$ to $1M$; Figure \ref{fig:fig8}(d) shows that the number of graph edges ranged from $15K$ to $20M$. Figures \ref{fig:fig8}(a), \ref{fig:fig8}(b) and \ref{fig:fig8}(c) have the same number of points and the same parameter of interest, what makes it possible to join them and see what the performance in seconds of Figure \ref{fig:fig8}(b) corresponds to in terms of graph size and, also, to verify empirically that the SCN complexity cost is linear with factor $f$.

The comparison of the methods, in absolute numbers (seconds) was favorable to the Graph-Tree as demonstrated in Figure \ref{fig:fig8}(a). Analytically speaking the Graph-Tree is favored by two facts; first, the number of {\it external edges} only rises to a fraction of the number of graph nodes. Second, even if the graph size increases, a proper partitioning scheme can make the number of {\it external edges} grow slower than the growth of the graph size.\\

\begin{figure}[htb]
\centering
\includegraphics[width=0.5\textwidth]{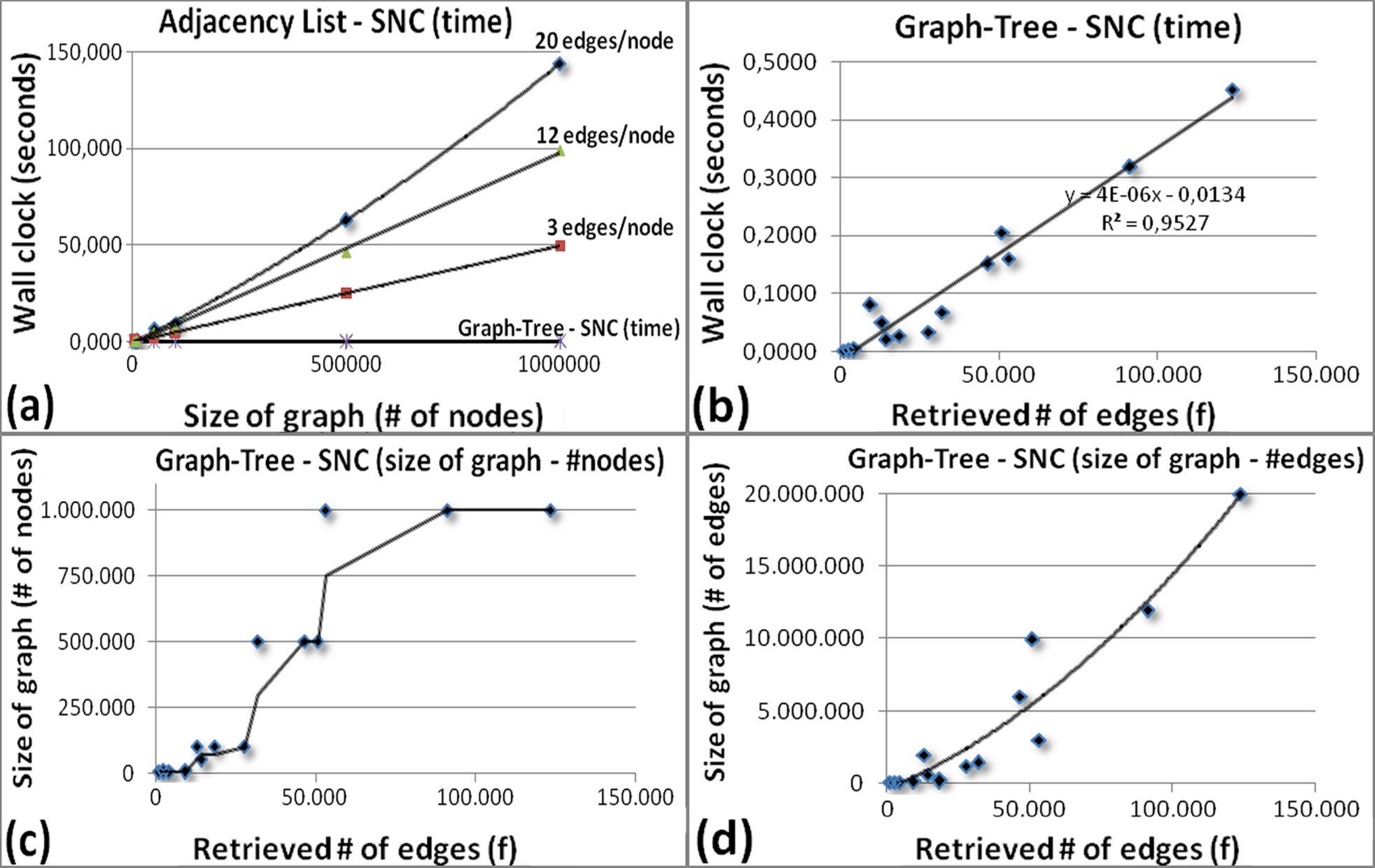}
\caption{Performance of SuperNodes Connectivity computation - 18 graphs (5K, 10K, 50K, 100K, 500K, 1M nodes) $\times$ (3, 12, 20) edges per node. (a) Adjacency list wall clock time for average degrees of 3, 12 and 20 edges per node, compared to Graph-Tree average time for several configurations of hierarchical partitioning and graph size. (b) Graph-Tree wall clock time for parameter $f$ (retrieved/expected number of edges between SuperNodes) -- linear complexity on $f$. (c) Size (number of nodes) of the graphs used for the measures showed in (b). (d) Size (number of edges) of the graphs used for the measures showed in (b).}
\label{fig:fig8}
\end{figure}

\noindent{\textbf{Performance on GNC Computation}}\\
\noindent{For GNC, our first observation is that the performance of the database was almost two orders of magnitude worse than the other two methods; its performance degrades heavily with the increase in the number of graph nodes and edges. The weak performance of the commodity database, once more, is due to the nested queries over the large volumes of information. It is explained by the inadequacy of the relational data model in calculating the GNC, which involves data crossing and tracking of the groups and subgroups to which the graph nodes pertain.}

\begin{figure}[htb]
\centering
\includegraphics[width=0.5\textwidth]{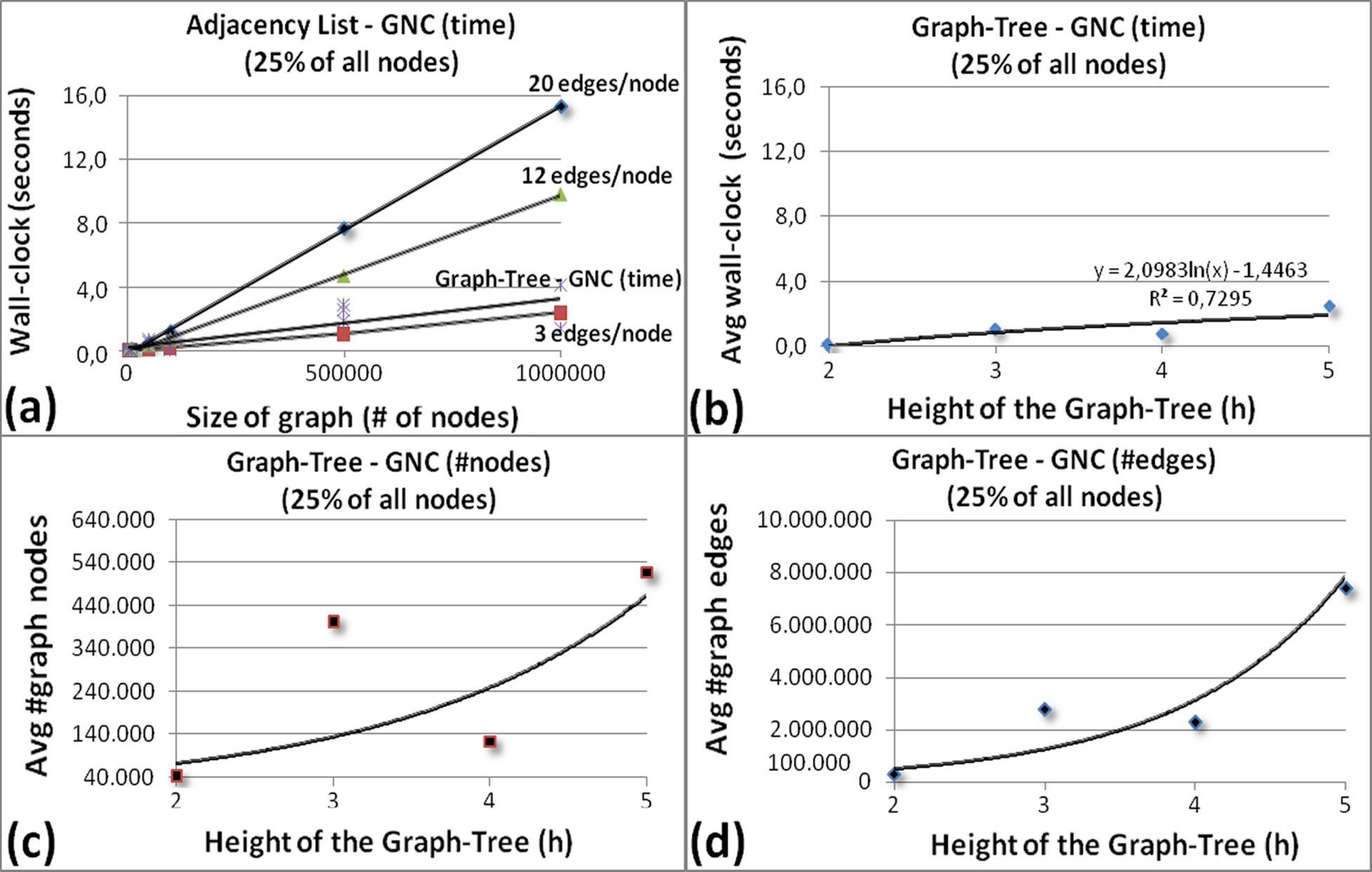}
\caption{Performance for Graph Nodes Connectivity - 18 graphs (5K, 10K, 50K, 100K, 500K, 1M nodes) $\times$ (3, 12, 20) edges per node, computed for $25\%$ of all the graph nodes. (a) Adjacency list wall clock time for average degrees of 3, 12 and 20 edges per node, compared to Graph-Tree average time for several configurations of hierarchical partitioning and graph size. (b) Graph-Tree wall clock time for parameter $h$, height of the Graph-Tree - logarithm complexity in accordance to the height of the tree. (c) Average size (number of nodes) of the graphs used for the measures showed in (b). (d) Average size (number of edges) of the graphs used for the measures showed in (b).}
\label{fig:fig9}
\end{figure}

Again here, as we see in Figure \ref{fig:fig9}(a), the adjacency list performance goes with the graph-size, having a reasonable performance. Actually, its performance is slightly better than the Graph-Tree for small edge degrees at the expense of larger memory demands. The strong point of the Graph-Tree is that although it is influenced by the graph size, as analytically predicted, its performance is not directly determined by this factor, but by the height ($h$) at which a given graph node of interest lies on -- a logarithmically increasing factor.

Just as for the SNC analysis, the different natures of the techniques ask for specific testing configurations. While Figure \ref{fig:fig9}(a) is ruled by the number of graph nodes and edges, Figures \ref{fig:fig9}(b), \ref{fig:fig9}(c) and \ref{fig:fig9}(d) are linked by the same number of points and by the same parameter of interest $h$. The joint of these three figures demonstrate the logarithmic characteristic of the Graph-Tree in numbers; while the curve in Figure \ref{fig:fig9}(b) range from $0.001$ second to nearly $3.5$ second, Figures \ref{fig:fig9}(c) and \ref{fig:fig9}(d) show that the average data used during the time experiment ranged from $40K$ to $540K$ nodes and from $100K$ to $8M$ edges. We note that average was used because it is not feasible to calculate all the possible hierarchical partitionings given by the combinations of number of levels $h$ and number of partitions per level for each of the $18$ graphs, therefore we have uniformly chosen random possibilities and combined their results with average; nevertheless all the possible graph sizes were used.

The GNC computational cost of the Graph-Tree grants a natural scalability potential that is not dictated by the graph size -- this is a demand for today's applications. By using a tree-like graph storage that supports GNC computation, it becomes possible to use all the classical graph algorithms without having the entire graph on memory, providing large scale possibilities.\\

\section{{\CEPS} Accuracy}
\label{sec:ceps_accuracy}

\noindent{In this section, we evaluate the accuracy of the {\CEPS} solution, rather than comparing it to other orthogonal approaches. We are interested in evaluating whether its algorithm captures the most relevant subgraph, given a desired budget size.}

The goodness score of an induced subgraph is measured through a simple question: ``how much importance is captured by the graph nodes that comprehend an induced subgraph CP?''. We refer to this measure as the ``{\em importance node ratio}'', or $IRatio$. Given a query set $\QMySet$ of nodes, a subgraph $G'$ and a connection subgraph $CP$, the $IRatio$ refers to the coefficient between the goodness score w.r.t. the induced connection subgraph $CP$ and the goodness score w.r.t. the entire subgraph $G'$. This computation assumes, as discussed in Section \ref{subsec:goodness}, that the goodness score used by {\CEPS} is accurate on its goal to measure the goodness of a graph. $IRatio$ is computed as follows:

\begin{equation}
\label{eq:iratio}
IRatio = \frac{{\sum\limits_{j \in CP} {r(\QMySet,j)} }}{{\sum\limits_{j \in G'} {r(\QMySet,j)} }}
\end{equation}

We use the $IRatio$ to evaluate the quality of {\CEPS}. In our experiments, we apply the {\CEPS} algorithm to the leaf communities of the DBLP dataset, each community containing around $500$ nodes. Figure \ref{fig:fig10} shows the average $IRatio$ versus size of subgraph (budget); the curves indicate the different query set sizes of our experiments. One can see that a relatively small connection subgraph (with 20 to 30 nodes) can capture most of the important nodes (accounting for $>$80\% of the total importance). This result shows that the {\CEPS} algorithm sticks to the essence of the original graph as much as possible, while considering the budget size limit.

\begin{figure}[htb]
\centering
\includegraphics[width=0.49\textwidth]{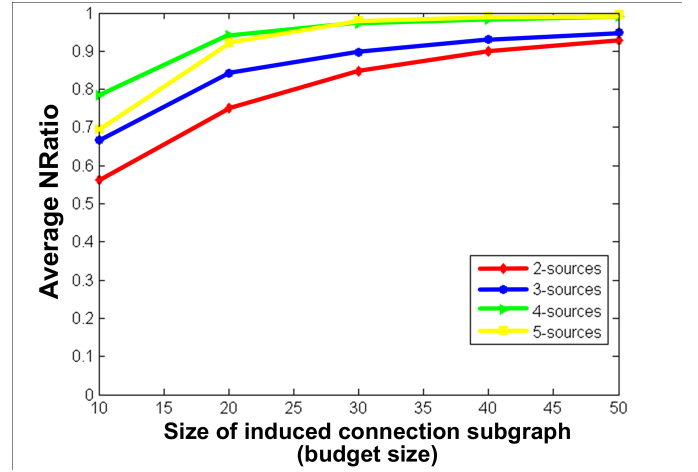}
\caption{Quality of the {\CEPS} summarization. The average ratio of important nodes in the induced {\CEPS} subgraph, varying the budget size and the number of query nodes (sources).}
\label{fig:fig10}
\end{figure}

\section{Proof of Concept: GMine Visual Environment}
\label{sec:SuperGraph_Visualization_in_GMine}

\noindent{Here we introduce the GMine system that, using the Graph-Tree structure, materializes SuperGraphs for visual inspection. Due to space limitations, it is not possible to show all the features of the system, so we have made it available at \url{http://www.cs.cmu.edu/~junio}. The dataset we use in this paper define authorship graphs deriving from publication data; each graph node represents an author and each edge denotes a co-authoring relationship.}\\

\noindent{{\bf DBLP Dataset}}\\

\noindent{Here we present the functionalities of GMine over a larger dataset. We use the Digital Bibliography \& Library Project (DBLP), a database of Computer Science publications. DBLP defines an authorship graph with $315,688$ nodes (authors) and $1,659,853$ edges (co-authorings). We use GMine to automatically create a recursive partitioning of DBLP according to the k-way partitioning (METIS). The partitioning has $5$ hierarchy levels, each with $5$ partitions. The dataset, thus, is broken into $5^{(5-1)}$, or $625$, communities with an average of nearly $500$ nodes per community. For this dataset, such partitioning generates communities anchored on highly collaborative authors and, roughly, on similar research themes.}

\begin{figure}[htb]
        \centering         
\includegraphics[width=0.49\textwidth]{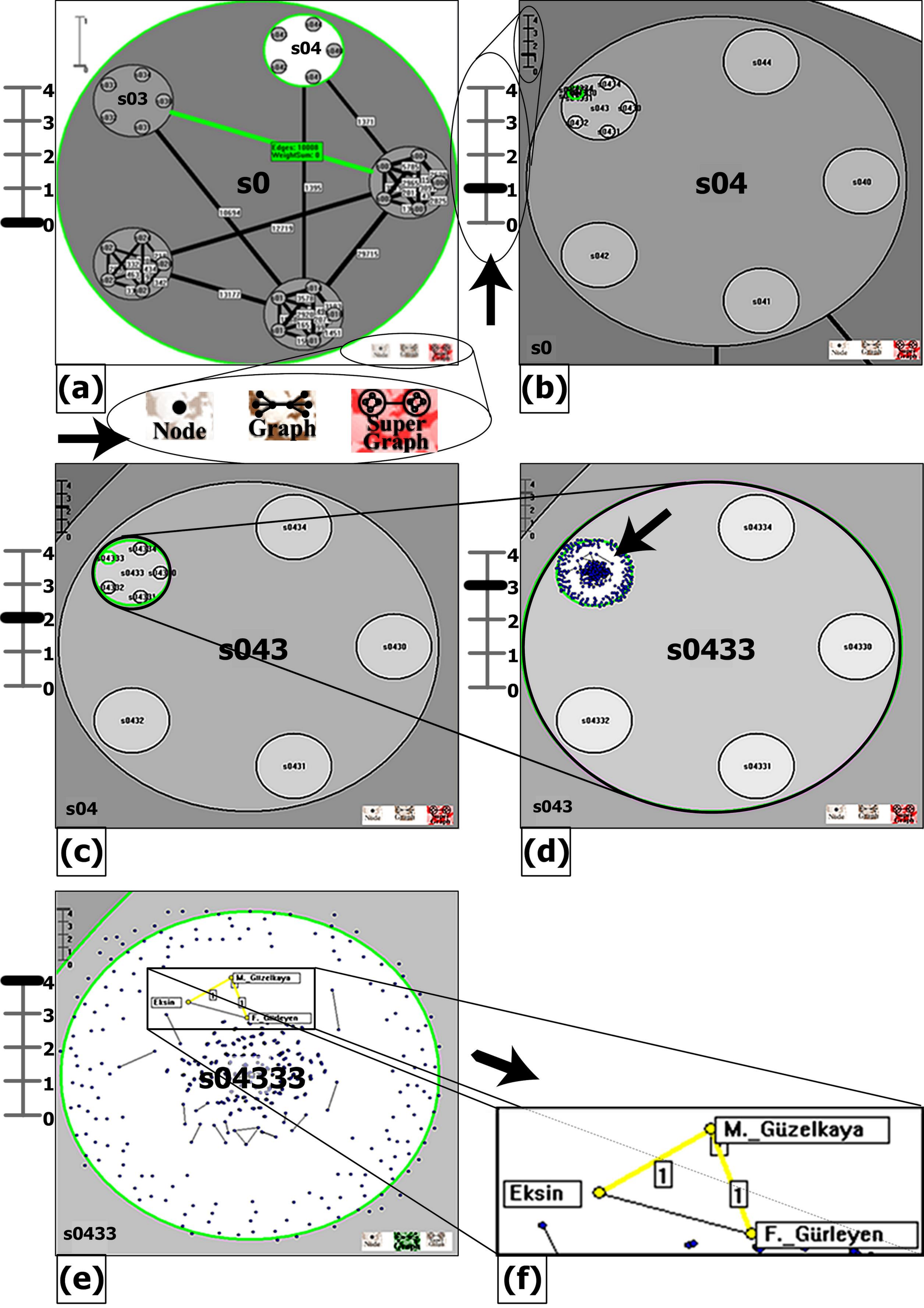}
        \caption{(a) Overview of DBLP dataset and highlight of the abstraction-control. (b) Focus on community {\it s04} and highlight of levels-selection control. (c) Focus on community {\it s043} and highlight of community {\it s0433}. (d) Zoom-in view of community {\it s0433} and the expansion of community subgraph {\it s04333}. (e) Inspection of community subgraph {\it s04333}, and highlight of one of its isolated sub-communities. (f) The sub-community embraces authors M. G\"uzelkaya, Eksin, and F. G\"urleyen.}
        \label{fig:fig6}
\end{figure}

\subsection{Visualization and Interaction}

\noindent{Figure \ref{fig:fig6} presents a navigation sequence over DBLP. In Figure \ref{fig:fig6}(a), it is possible to see the $5$ first-level partitions. By observing the SuperNodes connectivity (SuperEdges), it is possible to see that there are $3$ first-level communities highly connected one to each other, and that each of them also has their $5$ sub-communities highly inter-connected. The other $2$ first-level communities are relatively isolated, just similarly to their inner sub-communities. It is possible to conclude that the $3$ first-level highly connected communities hold long term collaborating authors, while the other $2$ -- {\it s03} and {\it s04} -- hold less productive casual authors who seldom interact with each other, or with authors from other communities.}

In Figure \ref{fig:fig6}(a) we highlight the {\em abstraction-control} of GMine (arrow below the figure), which allows to set the control to one of three abstraction entities: the individual graph nodes, the subgraphs at the leaves, or the SuperNodes of the SuperGraph. Figure \ref{fig:fig6}(b)
focuses on community {\it s04} and also shows (arrow at the left) the {\em  levels-selector control} of GMine, which permits the navigation through the levels of the hierarchy. In Figures \ref{fig:fig6}(c) and \ref{fig:fig6}(d) we go deeper into SuperNode {\it s04}, focusing on community {\it s043} and, further, on community {\it s0433}. Figure \ref{fig:fig6}(d) also shows that a leaf community of SuperNode {\it s0433} was loaded from disk (see arrow) under request of the user. In Figure \ref{fig:fig6}(e), community {\it s04333} is then presented with details about the nodes and edges of the correspondent subgraph. At this point, we have reached the deepest level of the SuperGraph. The detailed annotations on community {\it s04333} characterize its parent community {\it s04}, which contains mostly isolated nodes at the surroundings, and a few small subgraphs at the center. In Figure \ref{fig:fig6}(f), we focus on one of the subgraphs, which embodies 3 authors M. G\"uzelkaya, Eksin, and F. G\"urleyen. With the aid of the Graph Node Calculus (Section \ref{subsec:gNodesConnectivity}), we could retrieve their connections to the rest of the graph. We verified that none of them has additional co-authorings and, thus, their subgraph corresponds to their unique publication, dated from 2001.

\begin{figure}[htb]
        \centering
\includegraphics[width=0.49\textwidth]{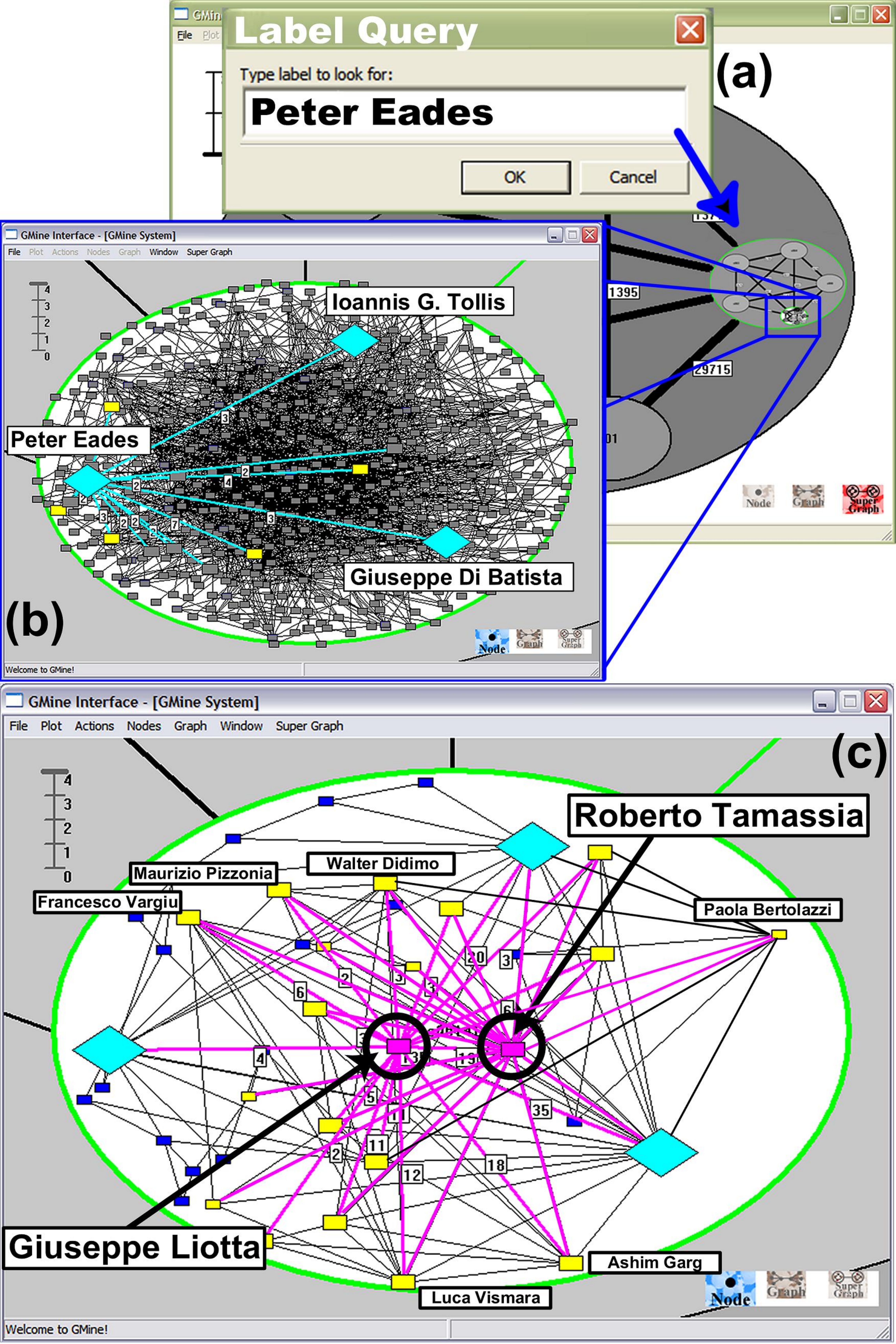}
        \caption{{\CEPS} illustration. (a) Label query for author Peter Eades indicates where the correspondent graph node is. (b) $500$ nodes community with highlighted authors Peter Eades, Ioannis G. Tollis and Giuseppe Di Batista. (c) $40$-nodes {\CEPS} presents a solid graph research community with highlighted authors Roberto Tamassia and Giuseppe Liotta, among others.}
        \label{fig:fig7}
\end{figure}

GMine also supports {\em label search} via hashing from the graph nodes to the SuperNodes of the Graph-Tree. In Figure \ref{fig:fig7}(a) we perform a {\em label search} for prominent graph analysis researcher Peter Eades; GMine takes us to the correspondent community indicated by the arrow. This subgraph, presented in Figure \ref{fig:fig7}(b) has around $500$ nodes cluttered in a limited space. At this point we can apply the {\CEPS} summarization to concentrate on a group of the most interesting graph nodes. As input, we pick authors Peter Eades, Ioannis G. Tollis and Giuseppe Di Battista, defining budget size of $40$ as the limit for the induced subgraph. Figure \ref{fig:fig7}(c) presents the final configuration, in which each graph node is connected to every other by a path smaller or equal $3$. The induced graph delineates a collaboration network where the query authors are cornerstone. Interestingly, the subgraph reveals two center-piece authors, Roberto Tamassia and Giuseppe Liotta, as central connections for the summarization subgraph. The entire subgraph presents one of the most remarkable graph research communities in the literature. This is only the main community for author Peter Eades; by calculating the Graph Node Connectivity, we verified that he has other $29$ co-authors from other partitions (communities) in this snapshot of DBLP.\\

\section{Conclusions}
\label{sec:Conclusions}

\noindent{We presented GMine, a system for large graphs visual analysis. The framework that supports GMine can process large graphs with hundreds of thousands of nodes using hierarchical graph partitioning and interactive summarization. Contributions include scalability via an innovative formalization for graph hierarchies aimed at graph processing and representation, an innovative connection subgraph extraction algorithm, and a proof-of-concept presentation of large graphs.}

As future research, we foresee the Graph-Tree purely designed for disk access, probably having its design oriented to SuperEdges; algorithms over the Graph-Tree for large graphs computation, benefiting from its plenary representation with GNC and SNC; the advancement of the SuperGraph abstraction for dealing with SuperNodes as if they were sole graph nodes, with specific properties reflecting their coverage; and the use of the GMine framework along with state-of-the-art layout techniques both for graphs and graph hierarchies, this last application in demand for systematic user evaluation.\\

\begin{spacing}{0.8}
{\footnotesize
\noindent{{\bf Acknowledgments}}\\
\noindent{This work was partly supported by Microsoft Research, FAPESP (S\~ao Paulo State Research Foundation), CAPES (Brazilian Committee for Graduate Studies), CNPq (Brazilian National Research Foundation), the National Science Foundation under Grants IIS-0209107, SENSOR-0329549 and IIS-0534205, the Army Research Laboratory under CAN W911NF-09-2-0053, and DARPA under CAN W911NF-11-C-0200. This work was also partly supported by the Pennsylvania Infrastructure Technology Alliance (PITA) and by donations from Intel, NTT and Hewlett-Packard. Any opinions, findings, and conclusions or recommendations expressed in this material are those of the authors and do not necessarily reflect the views of any of the funding institutions.}
}
\end{spacing}


{\ }

\begin{spacing}{0.95}

\begin{wrapfigure}{L}{10mm}
\vspace{-11pt}
\includegraphics[width=11mm]{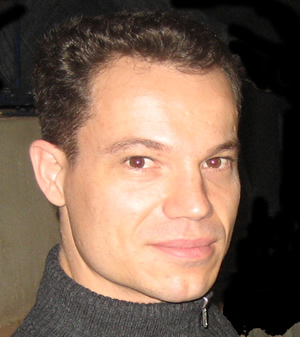}
\end{wrapfigure}

{\scriptsize
\noindent{{\bf \footnotesize Jose F. Rodrigues Jr.} is a Professor at University of S\~ao Paulo, Brazil. He received his Ph.D. from this same university, part of which was carried out at Carnegie Mellon University in 2007. Jose Fernando is a regular reviewer of major conferences in his field having contributed with publications in IEEE and ACM journals and conferences. His topics of research include data analysis, content-based data retrieval and visualization.}\\

\begin{wrapfigure}{L}{10mm}
\vspace{-11pt}
\includegraphics[width=11mm]{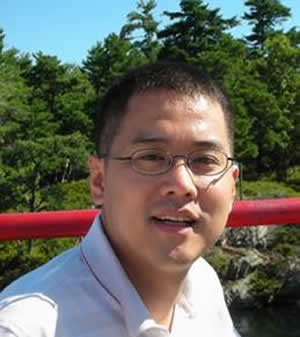}
\end{wrapfigure}

\noindent{{\bf \footnotesize Hanghang Tong} is a researcher at IBM T.J. Watson Research Center. He received his Ph.D. from the School of Computer Science, Carnegie Mellon University in 2009. Dr. Tong has received two "best paper" awards, pulished 40 papers, and filed eight patents. His research interests include data mining for graphs and multimedia.}\\

\begin{wrapfigure}{L}{10mm}
\vspace{-11pt}
\includegraphics[width=11mm]{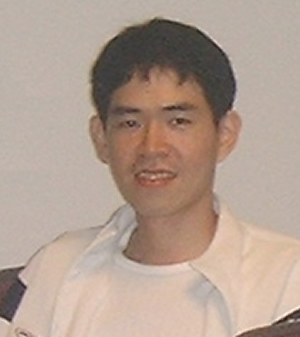}
\end{wrapfigure}

\noindent{{\bf \footnotesize Jia-Yu Pan} is a software engineer at Google Inc., USA, working on anomaly detection and its applications. He received his Ph.D. from Carnegie Mellon University, and has received three "best paper" awards. His research interests include anomaly detection, data mining, web services, and cloud computing.}\\

\begin{wrapfigure}{L}{10mm}
\vspace{-11pt}
\includegraphics[width=11mm]{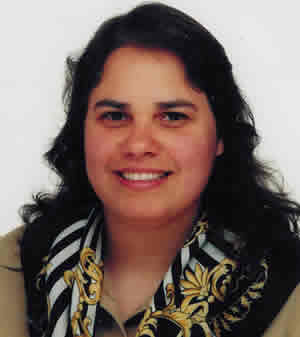}
\end{wrapfigure}

\noindent{{\bf \footnotesize Agma J. M. Traina} is a Professor at University of S\~ao Paulo having advised so far 32 graduate students, with over a hundred publications in major journals and conferences. Her research interests include multidimensional indexing methods, information visualization, retrieval by content, image processing and mining.}\\

\begin{wrapfigure}{L}{10mm}
\vspace{-11pt}
\includegraphics[width=11mm]{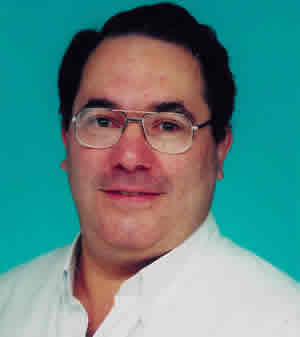}
\end{wrapfigure}

\noindent{{\bf \footnotesize Caetano Traina Jr.} is a Professor at University of S\~ao Paulo, being an active researcher in his field with over 200 referred papers, several awards and a large history of supervising graduate and undergraduate students. His research interests include database design, indexing methods, similarity queries and data mining.}\\

\begin{wrapfigure}{L}{10mm}
\vspace{-11pt}
\includegraphics[width=11mm]{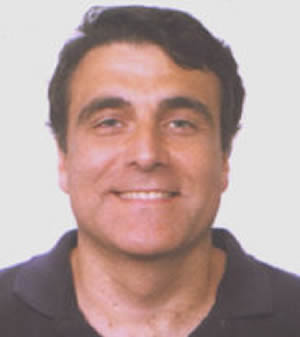}
\end{wrapfigure}

\noindent{{\bf \footnotesize Christos Faloutsos} is a Professor at Carnegie Mellon University. He has received the Research Contributions Award in ICDM 2006, the SIGKDD Innovations Award (2010), and seventeen ``best paper'' awards. He has published over 200 refereed articles, and 11 book chapters. His research interests include data mining for graphs and streams, database performance, and indexing for multimedia data.}

}\end{spacing}

\end{document}